\documentclass{appolb}
\usepackage{graphicx}
\newcommand{\req}[1]{(\ref{#1})}
\newcommand{\muF}{\mu_{F}}

\newcommand{\sla}{\hspace*{-0.10cm}/}
\def\phiDA{\phi}

\begin{document}
\title{Deeply-virtual and photoproduction of mesons
at higher-order and higher-twist 
\thanks{Presented at XXIX Cracow Epiphany Conference on Physics at the
Electron-Ion Collider and Future Facilities, Cracow, Poland, 16-19 January 2023}%
}
\author{ K. Passek-K.   
\address{Division of Theoretical Physics, Rudjer Bo\v{s}kovi\'{c} Institute,
Zagreb, Croatia }
\\[3mm]
}
\maketitle
\begin{abstract}
Both deeply-virtual and photoproduction of mesons offer promising access
to generalized parton distributions and complementary description 
of different kinematical regions. 
The higher-order contributions offer stabilizing effect with respect 
to the dependence on renormalization scales, 
while higher-twist effects have been identified as especially important 
in the case of the production of pseudo-scalar mesons. 
This was confirmed by recent evaluation of the complete twist-3 contribution 
to $\pi$ and $\eta$/$\eta'$ photoproduction 
and its confrontation with experimental data.
%
\end{abstract}
  
%
%
\section{Introduction}

Historically, most of the information about 
the high-energy nucleon structure came 
from the deeply inelastic scattering (DIS). 
From DIS data one extracts the parton distribution functions (PDFs) 
being the probabilities that a certain parton is found in a nucleon 
with certain longitudinal momentum fraction
of the nucleon momentum.
Through PDFs the one dimensional 
structure of the nucleon is thus revealed.
The hard-exclusive processes
offer insight into the transverse distribution of partons
and corresponding
generalized parton distributions (GPDs) give access 
to nucleon 3D structure. 
GPDs are functions of three variables: 
$x$, the parton's ''average'' longitudinal momentum fraction,
$\xi$, the longitudinal momentum transfer (skewness parameter), 
and $t$, momentum transfer squared,
while their evolution with energy is encapsulated in
the dependence on the factorization scale. 
At leading twist-2 there are eight quark GPDs and eight gluon GPDs 
classified according to different quantum numbers (parity, chirality),
as well as different GPDs for different quark flavours. 
To reveal their form is thus not an easy task 
and information from several processes have to be combined.

For the description of the hard exclusive processes
one employs the handbag mechanism in which
only one quark from the incoming nucleon
and one from the outgoing nucleon participate 
in the hard subprocess while all other partons
are spectators.
The simplest and well-investigated process to which 
this approach has been applied 
is Compton scattering $\gamma^{(*)} N \rightarrow \gamma N$,
while meson electroproduction $\gamma^{(*)} N \rightarrow M N'$
represents the natural extension and offers access to quark flavours. 
A prerequisite for the handbag mechanism is the presence 
of at least one large scale, which allows for the use of 
perturbative expansion in the strong coupling constant 
and the power, i.e., twist, expansion.
Two kinematic regions have been extensively studied: 
the deeply virtual (DV) region, where the virtuality $Q^2$ 
of the incoming photon is large, and the momentum transfer $(-t)$ 
from the incoming to the outgoing nucleon is small; 
and the wide-angle (WA) region, where $(-t)$, $(-u)$, and $s$ 
are all large, while $Q^2$ is smaller than $(-t)$ 
($Q^2=0$ in the case of photoproduction).
Factorization proofs exist for all orders 
for DV Compton scattering (DVCS) \cite{Collins:1998be} 
and DV meson production (DVMP) \cite{collins}, 
with the process amplitudes factorizing into hard perturbatively calculable 
subprocess amplitudes and GPDs that encapsulate the soft hadron-parton 
transitions and the hadron structure.
However, general factorization proofs are still lacking for WA processes, 
although it has been shown that factorization holds to next-to-leading order 
in the strong coupling for WA Compton scattering (WACS) \cite{rad98,DFJK1} 
and to leading order for WA meson production (WAMP) \cite{huang00}.
It is argued that in the symmetric frame where skewness is zero, 
the amplitudes can be represented as a product of subprocess 
amplitudes and form factors that represent $1/x$ moments of GPDs at zero-skewness.

Both DVCS and  WACS were widely investigated in the last decades 
and the handbag factorization achieved a good description 
of the experimental data.
The leading twist-2 description of DV vector meson production 
only considers the contributions of longitudinally polarized photons, 
specifically 
$\gamma^{*}_L N \rightarrow V_L N'$. 
This description has been observed to be in relatively 
good agreement with the current experimental data 
(see \cite{Aaron:2009xp, Adolph:2012ht} and the references therein). 
However, there is still a lack of systematic separation between 
longitudinal and transverse experimental data.
The contributions of transversely polarized photons 
$\gamma^{(*)}_T N \rightarrow V_{L,T} N'$
have also been investigated by 
including the twist-3 corrections to the meson state
\cite{Anikin:2002wg, Goloskokov:2013mba}.
On the other hand, 
the experimental data for DV pion production 
\cite{hermes,clas12,defurne,JeffersonLabHallA:2020dhq}
suggest the high significance of transversely polarized photons, 
which are not accounted for by the leading twist-2 
$\gamma^{*}_L N \rightarrow \pi N'$ contributions.
As in the vector meson case, a twist-3 calculation 
has been proposed, which 
incorporates twist-2 chiral-odd, 
i.e., transversity (parton helicity flip), GPDs 
and twist-3 pion corrections. 
The calculation including only the twist-3 2-body 
pion Fock component (Wandzura-Wilczek approximation) 
has already achieved a successful agreement with the data \cite{GK5}.
Experimental data for WA pion production 
\cite{anderson76,zhu05,clas-pi0} 
also indicate that the twist-2 contributions 
\cite{huang00} are not sufficient.
But unlike DVMP, the twist-3 contribution to pion photoproduction was found 
to vanish in the commonly used Wandzura-Wilczek approximation. 
In \cite{KPK18}, both 2- and 3-body twist-3 Fock components of $\pi_0$ 
were considered and successfully fitted to CLAS data \cite{clas-pi0}.
This work was extended to photoproduction of $\eta$ and $\eta'$ mesons
\cite{KPK21a} and 
WA electroproduction of $\pi^{\pm}, \pi^{0}$ \cite{KPK21}.
The application of the latter analytical results for the 
subprocess amplitudes to the DVMP subprocess amplitudes
is straightforward, and the phenomenological analysis is 
underway.

The DV and WA regions enable complementary access to GPDs 
at small and large $(-t)$, respectively.
A vast amount of experimental data needs to be confronted 
with reliable theoretical predictions, which should include 
higher-order perturbative predictions 
as well as higher-twist contributions. 
Here, we provide a brief overview of some recent developments.

\section{Deeply-virtual meson production at twist-2 and NLO}

The DVMP amplitude 
$\gamma^* N \to M N'$ 
can be expressed through, so-called, transition form factors
\begin{equation}
{}^{a}\mathcal{T}(\xi, t, Q^2) = 
\int_{-1}^1 \frac{{\rm d}x}{2\xi}\;
\int_0^1 {\rm d}u\;
T^{a} (x, \xi, u, \mu) 
\; F^{a} (x, \xi, t, \mu)
\; \phi(u,\mu)
\, 
\label{eq:tff}
\end{equation}
with $a$ denoting quark and gluon contributions, and
$u$ the longitudinal momentum fraction of the meson's parton.
The factorization scale $\mu$ separates the short-distance dynamics, 
represented by the subprocess amplitudes $T^{a}$, 
from the long-distance dynamics represented by the hadron wave functions: 
the GPD $F^{a}$ and the meson distribution amplitude (DA) $\phi$.

Transition form factors 
${}^{a}\mathcal{T}$ have the similar role in DVMP as
Compton form factors in DVCS,
but they additionally depend on meson DA,
i.e., meson structure, making the analysis
of the process both more challenging, 
as well as, potentially more rewarding.
In contrast to DVCS, DVMP enables easy access
to GPDs of different quark flavours,
and offers the natural distinction of 
GPDs of different parity: 
at twist-2 chiral-even GPDs $H^q$, $E^q$ contribute to the production
of longitudinally polarized vector mesons ($V_L$) 
and scalar ($S$) mesons,
while $\widetilde{H}^q$ and $\widetilde{E}^q$ 
appear in the production of pseudoscalar ($P$) 
and axial-vector ($A_L$) mesons.
Moreover, the contribution of gluon GPDs $H^g$, $E^g$
($\widetilde{H}^g$, $\widetilde{E}^g$) 
to the production of neutral $V_L$ ($A_L$) mesons 
is more significant since, unlike in DVCS, they contribute 
already at the leading order. 
Therefore, their form is phenomenologically more accessible.

The twist-2 DVMP subprocess amplitudes
$\gamma^*_L q \to (q \bar{q}) q$ 
and
$\gamma^*_L g \to  (q \bar{q}) g$ 
are calculated perturbatively 
order by order in the strong coupling constant
\begin{equation}
T^a(x, \xi, u, \mu)
=
\frac{\alpha_s(\mu_R)}{4\pi}
T^{a(1)}(x, \xi, u)
+
\frac{\alpha_s^2(\mu_R)}{(4\pi)^2}
T^{a(2)}(x, \xi, u, \mu_R,\mu)
+
\cdots
\end{equation}
and they have been
determined to next-to-leading order (NLO) 
for flavour non-singlet and singlet $P$ and $V_L$ mesons
\cite{Belitsky:2001nq, Ivanov:2004zv,Duplancic:2016bge},
as well as, for the (crossed) production of $S$ and $A_L$
mesons \cite{Muller:2013jur,Duplancic:2016bge}.
Predictions at finite order are inherently dependent 
on the renormalization scale $\mu_R$ and scheme, 
introducing additional theoretical uncertainty. 
Therefore, the inclusion of higher-order corrections is crucial 
to reduce this dependence and stabilize predictions. 
Although meson DAs ($\phi$) and GPDs ($F^a$) 
are intrinsically nonperturbative quantities, 
their evolution can be calculated perturbatively. 
The complete closed form is known to NLO order \cite{Belitsky:1998uk}, 
and more recently, NNLO contributions to the evolution kernels 
have been obtained \cite{Braun:2017cih}.

\begin{figure}
\centering
\includegraphics[width=\textwidth]{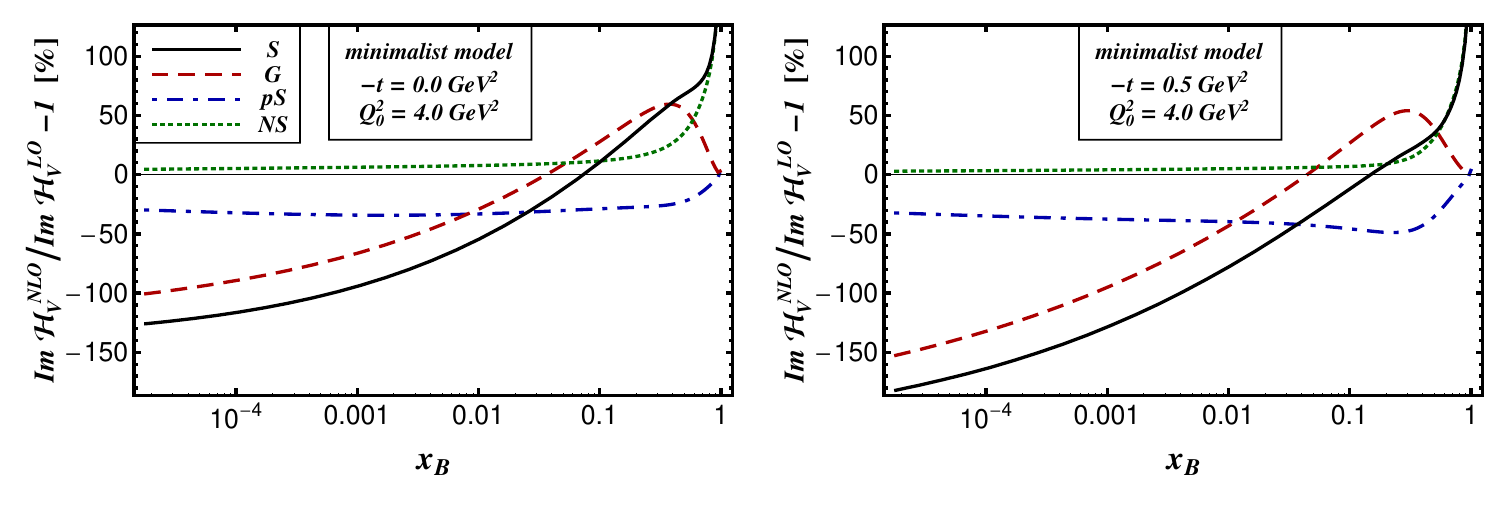}
\caption{\small
Relative NLO corrections to the imaginary part of the flavor singlet TFF
(solid) broken down to the
gluon (dashed), pure singlet quark  (dash-dotted) and `non-singlet' quark
(dotted) contributions (Ref. \cite{Muller:2013jur}).
}
\label{Fig-S-Im}
\end{figure}

\begin{figure}
\centering
\includegraphics[width=0.75\columnwidth]{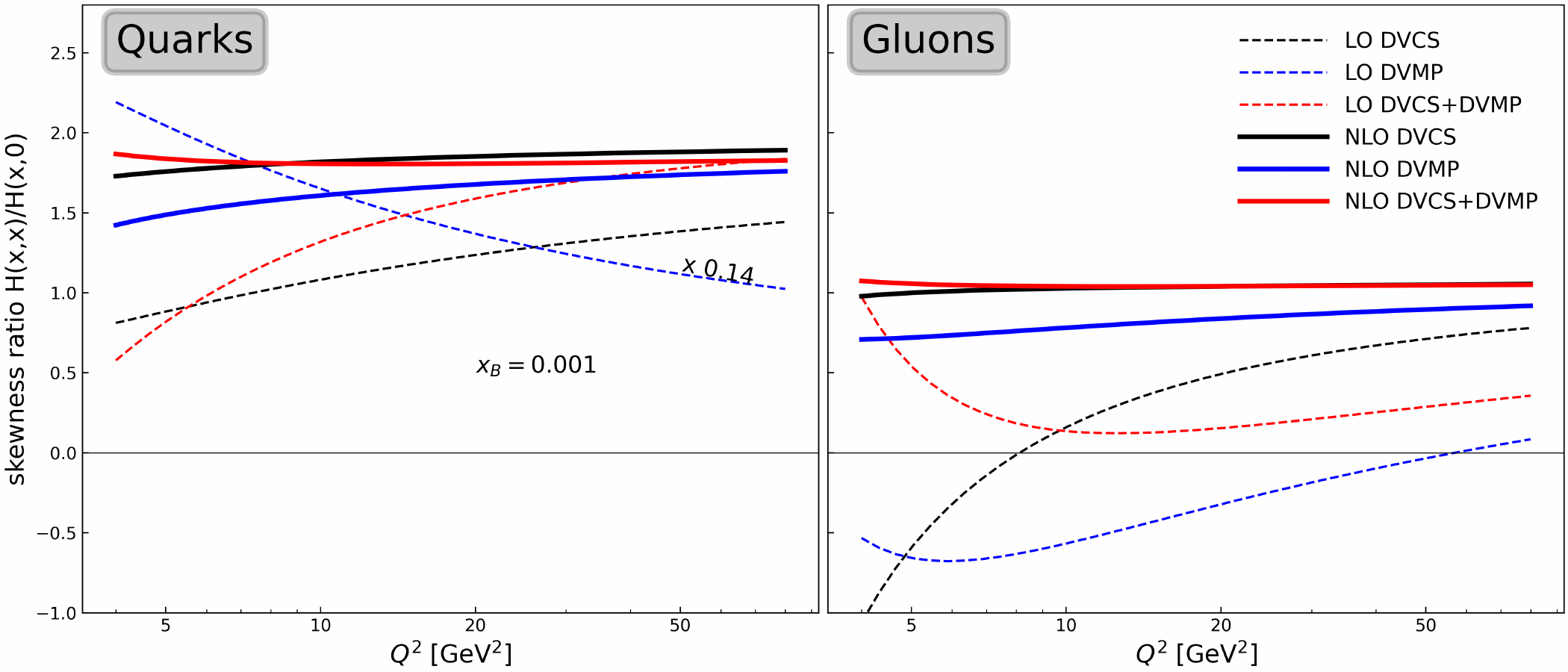}
\caption{Skewness ratio for GPD $H$ (preliminary K. Kumerički, Transversity 2022).}
\label{fig:ratio}
\end{figure}

The evolution is simpler to implement in the conformal
momentum representation.
Conformal moments are analogous to Mellin moments in
DIS and represent the moments with respect to 
the eigenfunctions of the leading order evolution
kernels, i.e., with respect to
Gegenbauer polynomials $C_n^{3/2}$ and $C_n^{5/2}$
for quarks and gluon, respectively.
The convolution over $x$ and $u$ in transition form factors
\req{eq:tff} is thus replaced by the summation over
conformal moments, and consequently the
series is summed using the Mellin-Barnes integral over 
complex conformal moment $j$ \cite{Muller:2013jur}
\begin{eqnarray}
 {}^{a}\mathcal{T}(\xi, t, Q^2) 
&=& \frac{1}{2i}\int_{c-i \infty}^{c+ i \infty}\!
 dj \left[i\pm 
 \left\{ {\tan \atop \cot} \right\}
 \left(\frac{\pi j}{2}\right) \right] \,\xi^{-j-1}
\nonumber \\
& &
\times 
\left[
T^a_{jk}(\mu)
\stackrel{k}{\otimes}
\phi_{k}(\mu)
\right]
F^a_j(\xi, t, \mu)
\, .
\label{eq:tff-conf}
\end{eqnarray}
This approach has been developed and extensively applied to DVCS
\cite{Kumericki:2007sa}, and then extended to DVMP. 
Regardless of whether one considers Compton or transition form factors 
in momentum fraction (\ref{eq:tff}) 
or conformal momentum space (\ref{eq:tff-conf}), 
complete deconvolution is impossible, and GPD access is only possible 
through different modelling approaches.
Dedicated software is now available: PARTONS \cite{Berthou:2015oaw} 
and Gepard \cite{gepard}
for analysis in momentum fraction and conformal momentum space, respectively.
The DV$V_L$P is included in the latter.

While there has been a lot of interest in the DVCS process, 
there are relatively few NLO phenomenological analyses of the DVMP process
\cite{Belitsky:2001nq,Diehl:2007hd,Muller:2013jur}, 
despite the availability of experimental data. 
The complete set of $x$ and $j$ space analytical results for 
all meson channels can be found in \cite{Muller:2013jur,Duplancic:2016bge}.
The numerical analysis performed there shows 
that NLO corrections are important and model-dependent 
(Fig.\ref{Fig-S-Im}).
The effects of LO GPD and DA evolution are significant
and for NLO calculations one should include also NLO evolution.
Gluon corrections play a significant role in small $\xi$ production of vector mesons, 
and there may be a need for resummation of the large 
logarithmic $\ln(1/\xi)$ terms
observed in both gluon evolution and gluon coefficient function. 
Finally, the choice of meson distribution amplitude is found to 
have a significant impact on the results.

Since GPDs are process-independent quantities,
the simultaneous description and global
fits of GPDs to DIS, DVCS and DVMP data
represent the next necessary step.
Through these one hopes to gain additional information
on the importance and stability of NLO predictions
and validity of different models.
Using the conformal momentum representation \cite{Muller:2013jur}
the first global fits on DIS, DVCS and DV$V_L$P 
small-x HERA collider data have been performed
at LO \cite{Meskauskas:2011aa}  
($\chi^2/n_\mathrm{d.o.f} \approx 2$),
and at NLO \cite{Lautenschlager} 
(Bayesian analysis).
The recent NLO analysis 
using corrected NLO analytical results \cite{Duplancic:2016bge}
and Gepard software 
shows promising agreement
of theory and experiment
($\chi^2/n_\mathrm{d.o.f}$ = 254.3/231)
and 
indicates that a global description of DVCS and DV$V_L$P 
is reachable at
NLO (Fig. \ref{fig:ratio}).

\section{Pseudoscalar meson production at higher-twist}

\begin{figure}
\centering
  \includegraphics[width=0.45\columnwidth]{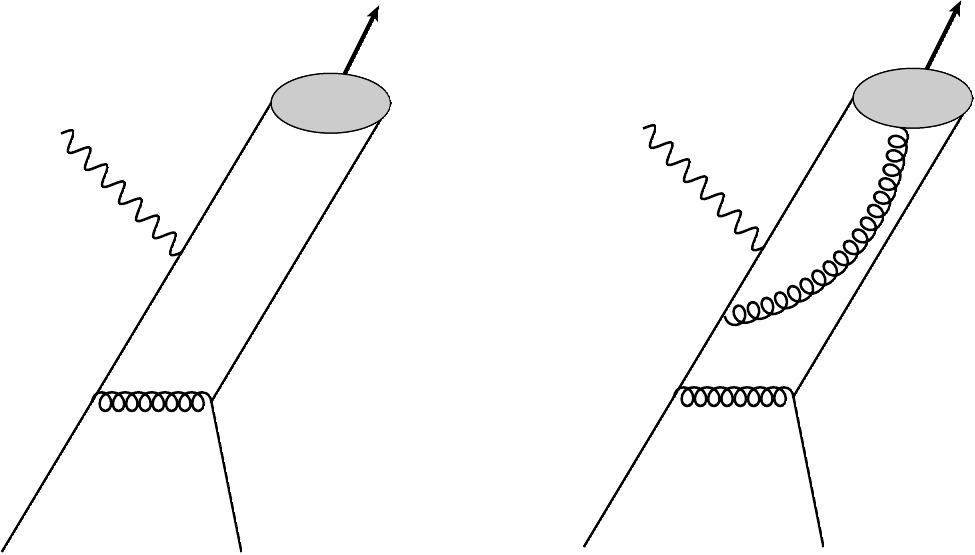} 
\caption{Generic diagrams for 2- and 3-body subprocess amplitudes.}
\label{fig:H}
\end{figure}

The twist-3 LO prediction for the electroproduction 
of the pseudoscalar meson $P$, 
which includes 2- and 3-body meson Fock states, 
was first calculated for the WA region \cite{KPK21}. 
Here, we review the findings and their confrontation with experimental 
data for photoproduction ($Q^2=0$). 
The analytical expressions obtained for the subprocess amplitudes 
can also be applied for the DV$P$P analysis ($t\to0$).

The helicity amplitudes for
$\gamma^{(*)} N \rightarrow P N'$
process in the WA angle region
can be expressed 
in terms of 
the subprocess amplitudes
${\cal H}$ 
multiplied by
the soft form factors, $R_i^{P}$ and $S_i^{P}$, which
represent 
of $1/x$-moments of zero-skewness GPDs 
$\int_0^1 \frac{dx}{x}\, F_i^{a}(x,t)$. 
The $R$-type form factors 
are related to the helicity non-flip GPDs
$H$, $\widetilde{H}$ and $E$. 
The $S$-type form factors 
are related to the helicity-flip or transversity GPDs $H_T$, $\bar{E}_T$
and $\widetilde{H}_T$
\footnote{The GPDs $\tilde{E}$ and $\tilde{E}_T$ and their
associated form factors decouple at zero skewness.}.

The amplitudes ${\cal H}$ 
correspond to the subprocesses
$\gamma^{(*)} q \rightarrow P q'$
and they are calculated using handbag diagrams
as the ones
depicted on Fig. \ref{fig:H}.
The meson $P$ is replaced by an appropriate
2- or 3-body Fock state.
The projector $\pi \rightarrow q\bar{q}$ 
contributes to the subprocess amplitudes
corresponding to the diagrams 
depicted on Fig. \ref{fig:H}a 
and its structure
is given by
\begin{equation}
{\cal P}_{2}^{P} \sim f_\pi
   \,\Big\{\gamma_5 \,p\!\sla\phiDA_\pi(u,\mu) 
                   + \mu_\pi(\mu) \gamma_5 \Big[\,\phiDA_{\pi p}(u,\mu)  
- [ \ldots ] \phiDA'_{\pi\sigma}(u,\mu)
+ [ \ldots ] \phiDA_{\pi\sigma}(u,\mu)\Big] \Big\}
\, .
\label{eq:2proj}
\end{equation}
The first term in \req{eq:2proj} corresponds to the twist-2 part,
while the twist-3 part is proportional to the chiral condensate
$\mu_\pi=m_\pi^2 /(m_u+m_d)\cong 2$ GeV (at the factorization scale $\muF=2$ GeV).
This parameter is large and although the twist-3 
cross-section for pion electroproduction is 
suppressed by $\mu_\pi^2/Q^2$ 
as compared to twist-2 cross section, 
for the range of $Q^2$ accessible in current
experiments the suppression factor is of order unity%
\footnote{
Twist-3 effects can also be generated by twist-3 GPDs. 
However, these are expected to be small and therefore
neglected.}%
.
The 3-body $\pi \rightarrow q \bar{q} g$ projector 
contributes to the amplitudes corresponding to 
Fig. \ref{fig:H}b, 
\\
\begin{equation} 
{\cal P}_{3}^{P} \sim
             f_{3\pi}(\mu) \gamma_5 [ \ldots ]\,
  \phiDA_{3\pi}(u_1,u_2,u_g, \mu)
\, .
\label{eq:3proj}
\end{equation} 
The helicity non-flip amplitudes are generated by twist-2, while
the helicity flip ones are of twist-3 origin.

In addition to twist-2 DA $\phiDA_\pi$ there are two 2-body twist-3 DAs,
$\phiDA_{\pi p}$ and $\phiDA_{\pi \sigma}$, and 3-body twist-3 DA 
$\phiDA_{3 \pi}$.
Twist-3 DAs are connected by equations of motion (EOMs).
By EOMs and DA symmetry properties, 
it is possible to express the twist-3 subprocess amplitudes 
in terms of only two twist-3 DAs, and combine 2- and 3-body contributions.
Applying EOMs also results in an inhomogeneous linear first-order differential 
equation, which can be used to determine $\phi_{\pi p}$ (and $\phi_{\pi\sigma}$) 
from a known 3-body DA $\phiDA_{3 \pi}$ \cite{braun90}%
\footnote{It is important to note that the same gauge must be used consistently 
for the constituent gluon in the $q\bar{q}g$ projector and EOMs.} 
.

In meson electroproduction, both transverse and longitudinal photons 
contribute to twist-2 subprocess amplitudes. 
As expected, the longitudinal contribution vanishes in the photoproduction 
limit 
, while in the DVMP limit 
only longitudinal photons contribute. 
The general structure of the twist-3 contributions for both transverse 
and longitudinal photons reads 
\begin{eqnarray}
{\cal H}^{P,tw3}
\label{eq:Htw3}
& = & {\cal H}^{P,tw3,q\bar{q}} +  {\cal H} ^{P, tw3,q\bar{q}g}
\nonumber \\
&=&
   \big({\cal H}^{P,\phi_{\pi p}} + \underbrace{{\cal H}^{P,\phi_{\pi 2}^{EOM}}\big)+
   \big( {\cal H}^{P,q\bar{q}g, C_F}} + {\cal H}^{P,q\bar{q}g, C_G} \big)
\nonumber \\
&=&
   {\cal H}^{P,\phi_{\pi p}} 
\; \;  + \quad \qquad {\cal H}^{P,\phi_{3\pi},C_F}
\qquad \quad + {\cal H}^{P,\phi_{3\pi},C_G}
\, ,
\end{eqnarray}
where ${\cal H}^{P,tw3,q\bar{q}}$ is the twist-3 2-body contribution 
proportional to the $C_F$ color factor, and ${\cal H}^{P,tw3,q\bar{q}g}$ 
is the twist-3 3-body contribution with $C_F$ and $C_G$ proportional parts. 
The $C_G$ part is gauge invariant, whereas for $C_F$ contributions, 
only the sum of 2- and 3-body parts is gauge invariant with respect to the choice 
of photon or virtual gluon gauge. 
EOMs are used to obtain this sum, as well as the complete twist-3 contribution 
expressed through only two twist-3 DAs, $\phi_{3\pi}$ and $\phi_{\pi p}$.
The twist-3 subprocess amplitude for longitudinal photons
vanishes both for photoproduction and DVMP.
One finds that for 
photoproduction 
${\cal H}^{P,\phi_{\pi p}}=0$ 
\cite{KPK18}.
For DVMP, ${\cal H}^{P,\phi_{\pi 2}^{EOM}}=0$, 
and while no end-point singularities are present for $t \ne 0$, 
they must be considered in the limit $t \to 0$ 
since for 
${\cal H}^{P,\phi_{\pi p}} \sim \int_0^1 \frac{du}{u} \phi_{P p}(u)$.
In \cite{GK5} the modified hard-scattering picture
has been used to regularize the 2-body contributions.
With the complete twist-3 contribution 
now available \cite{KPK21} 
the analysis in modified and collinear picture
is underway.

\begin{figure}
\centering
  \includegraphics[width=0.45\columnwidth]{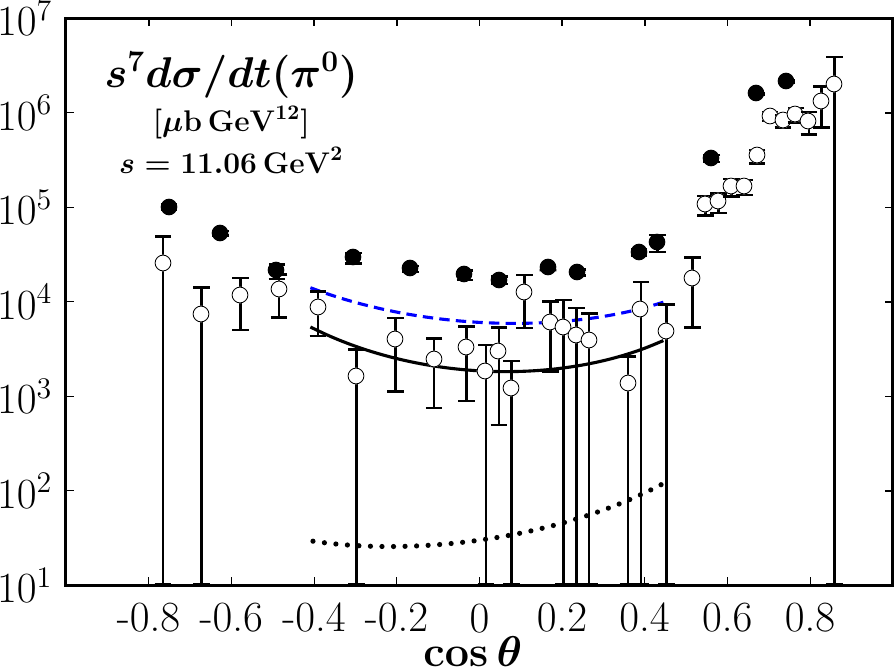} 
 \caption{ The cross section for $\pi^0$ photoproduction with twist-3 contributions.
    Solid (dotted) curve: full (twist-2) result.
    Dashed curve: full result with fixed renormalization and factorization scale.
    Data taken CLAS \cite{clas-pi0} (open circles) 
    and from SLAC \cite{anderson76} ($s=10.3$ GeV$^2$) 
   (Ref. \cite{KPK21}).}
\label{fig:photo}
\end{figure}

In \cite{KPK18} the cross-section for $\pi^0$ 
photoproduction has been fitted to \cite{clas-pi0} data.
The results are displayed in Fig. \ref{fig:photo}.
The twist-2 prediction lies well below the data. 
However, by including the twist-3 contributions 
one obtains reasonable agreement with the experiment.
Twist-3 is more important in the backward hemisphere ($\theta$ is c.m.s. scattering angle).
In \cite{KPK21}, the analysis was extended to $\pi^+$ and $\pi^-$, 
using only a few available experimental data \cite{anderson76,zhu05}. 
In \cite{KPK21a}, $\eta$ (preliminary GlueX data) and $\eta'$ photoproduction was studied. 
A similar behavior in photoproduction cross-sections was observed, 
except for $\eta'$, where the twist-2 contribution was significant, 
offering the possibility of determining the 2-gluon twist-2 DA.

For pion electroproduction there are
four partial cross sections. 
In \cite{KPK21} the theoretical predictions were given 
and the importance of the measurement was stressed.
Different combinations of form factors make 
it possible to extract transversity GPDs ($F^q_T$), 
which have a large $-t$ behavior that is important 
for parton tomography.

\begin{figure}
\centering
  \includegraphics[width=0.45\columnwidth]{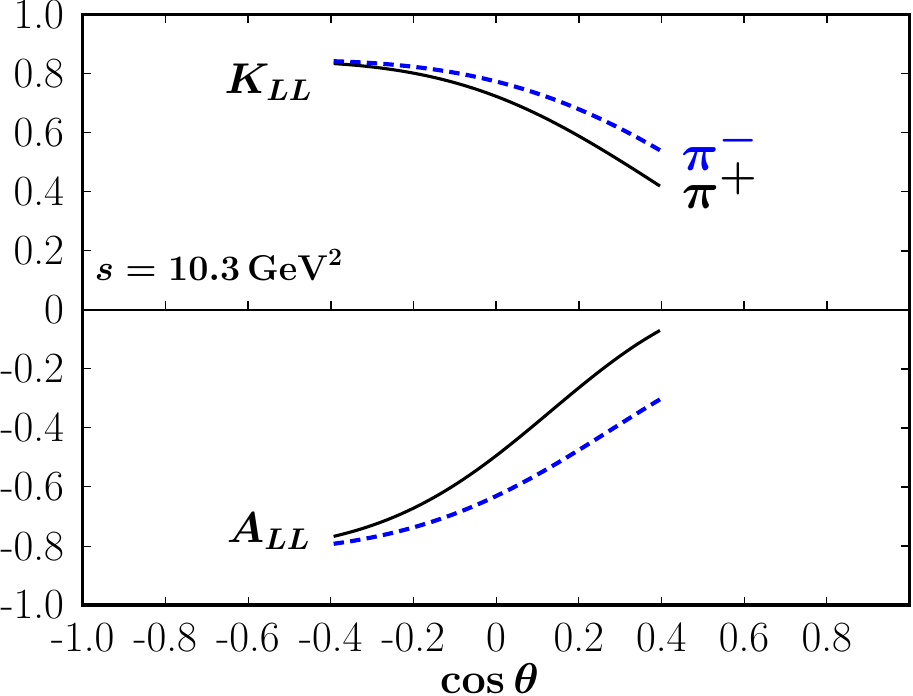} 
  \includegraphics[width=0.45\columnwidth]{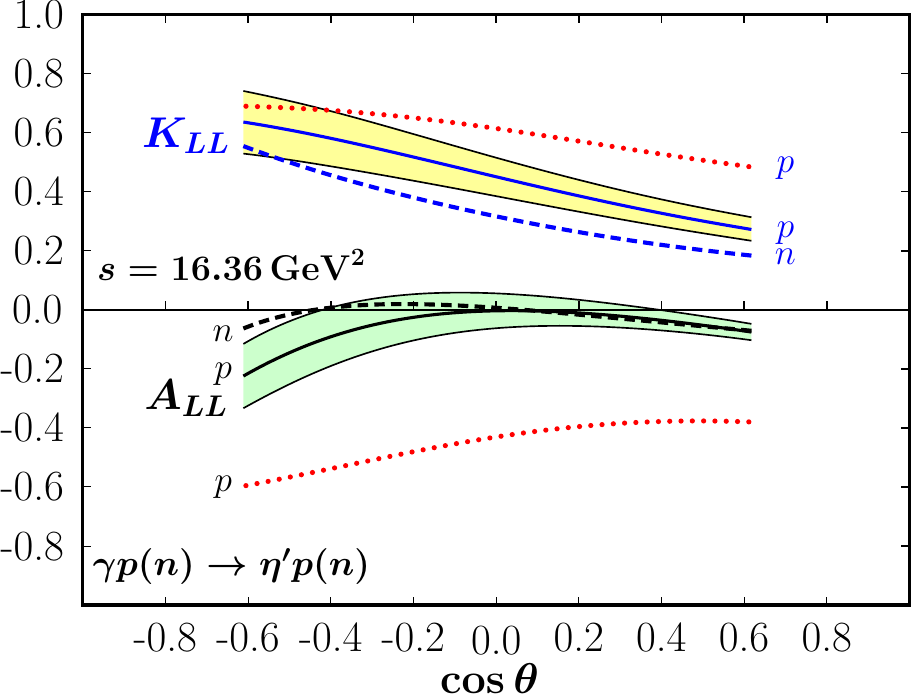}
\caption{Results for the helicity correlation parameters $A_{LL}$ and $K_{LL}$ 
  for $\pi^+$, $\pi^-$ and $\eta'$ photoproduction (Refs. \cite{KPK21,KPK21a}).}
\label{fig:AKLL}
\end{figure}

In meson photoproduction, spin-dependent observables such as 
the correlations of the helicities of the photon and either 
the incoming or outgoing nucleon, i.e., $A_{LL}$ and $K_{LL}$, 
offer additional insight that is less sensitive to particular parameters. 
It can be shown that $A_{LL}^{P,tw2} = K_{LL}^{P,tw2}$ 
and $A_{LL}^{P,tw3} = -K_{LL}^{P,tw3}$, 
indicating that the measurement of $A_{LL}$ and $K_{LL}$ offers a characteristic signature 
for the dominance of twist-2 or twist-3, similar to the role that the comparison 
of $\sigma_T$ and $\sigma_L$ has in DVMP. 
From Fig. \ref{fig:AKLL}, it is clear that our numerical results suggest the dominance 
of twist-3 for large $\theta$, while twist-2 increases in the forward direction.

\section{Summary and outlook}

Twist-2 NLO contributions to DVMP amplitudes are available 
and need to be compared with experimental data. 
The preliminary comparison of vector meson production to data seems 
satisfactory, but NLO corrections are significant, 
and the first DIS, DVCS, and DV$V_L$P fits have been performed. 
For pseudoscalar meson production, twist-3 contributions dominate, 
and a complete analysis of 2- and 3-body twist-3 contributions is ongoing. 
The available twist-2 NLO contributions should also be tested. 
It is important to note that the choice of meson distribution amplitude 
significantly affects the DVMP predictions. 
In WA photoproduction of $\pi$ mesons, the twist-2 analysis 
falls short by an order of magnitude. 
The complete twist-3 contribution has been included, and it was found that 
the meson's twist-3 contributions dominate for $\pi$s and $\eta$. 
Future experimental goals include the clear separation of longitudinally 
and transversely polarized photon contributions. 

{\it Acknowledgements}
This publication is supported
by the Croatian Science Foundation project IP-2019-04-9709,
by the EU Horizon 2020 research and innovation programme, STRONG-2020
project, grant agreement No 824093.




\end{document}